\documentclass[12pt,twoside]{article} 
\usepackage[T1]{fontenc} % font

\usepackage{amsthm, amsmath, amssymb, mathrsfs, upgreek, enumerate, mathtools, comment, natbib, graphicx, mathabx, booktabs, threeparttable, relsize, exscale, epstopdf, scrextend, multirow, xr-hyper, pdfpages, bm, caption, subcaption, tikz}
\usepackage[headheight=110pt,margin=1.25in]{geometry}
\usepackage[algoruled]{algorithm2e}
\usepackage[section]{placeins} % for \FloatBarrier command

\usepackage{fancyhdr, setspace} % headers and footers
\usepackage[pagebackref=false]{hyperref} % link colors
\hypersetup{
    colorlinks=true,
    citecolor=blue,
    filecolor=black,
    linkcolor=black,
    urlcolor=blue,
    bookmarksopen=true,
    pdfstartview=FitH
}

\newtheorem{theorem}{Theorem}
\newtheorem{assump}{Assumption}
\newenvironment{myassump}[2][]
  {\renewcommand\theassump{#2}\begin{assump}[#1]}
  {\end{assump}}
\newtheorem{proposition}{Proposition}

\newtheorem{secproposition}{Proposition}[section]

\newtheorem{seclemma}{Lemma}[section]

\theoremstyle{definition}
\newtheorem{definition}{Definition}

\newtheorem{example}{Example}

\newtheorem{remark}{Remark}

\newtheorem{secdefinition}{Definition}[section]

\def\Snospace~{\S{}}
\makeatletter
\def\thm@space@setup{
  \thm@preskip=15pt \thm@postskip=15pt % controls spacing before and
}
\makeatother

\def\indep{\perp\!\!\!\perp}
\newcommand{\argmax}{\operatornamewithlimits{argmax}}
\newcommand{\argmin}{\operatornamewithlimits{argmin}}

\newcommand{\E}{{\bf E}}
\newcommand{\R}{\mathbb{R}}

\newcommand{\N}{\mathcal{N}}
\newcommand{\U}{\mathcal{U}}

\newcommand{\C}{\mathcal{C}}
\newcommand{\prob}{{\bf P}}

\newcommand{\ind}{\bm{1}}
\providecommand{\abs}[1]{\lvert#1\rvert}

\let\emptyset\varnothing
\providecommand{\abs}[1]{\lvert#1\rvert}

\renewcommand{\qed}{\hfill \mbox{\raggedright \rule{0.08in}{0.08in}}} % black QED box
\renewenvironment{proof}[1][\proofname]{{\noindent\sc#1. }}{\qed\vspace{15pt}} % "proof" small caps

\title{\bf\sc Identifying Treatment and Spillover Effects Using Exposure Contrasts\thanks{I thank Yechan Park for insightful comments and seminar audiences at the 2025 Southern Economic Association Conference, Paris Econometrics Seminar, USC, and WUSTL for their feedback.}}

\author{Michael P.\ Leung\thanks{Department of Economics, University of California, Santa Cruz. E-mail: leungm@ucsc.edu.}}

\begin{document}
\maketitle
\onehalfspacing
 
\begin{abstract}

  {\sc Abstract.} To report spillover effects, a common approach is to regress outcomes on statistics summarizing neighbors' treatments. This paper studies nonparametric analogs of these estimands, which we refer to as exposure contrasts. We demonstrate that they may have the opposite sign of the unit-level effects of interest even under unconfoundedness. We then provide interpretable conditions on interference and the assignment mechanism under which exposure contrasts can be represented as convex averages of the unit-level effects, thereby avoiding sign reversals. These conditions encompass cluster-randomized trials, network experiments, and observational settings with peer effects in selection into treatment.

  \bigskip

  \noindent {\sc JEL Codes}: C21, C31, C57

  \noindent {\sc Keywords}: causal inference, identification, interference, peer effects
 
\end{abstract}

\newpage

%----------------------------------------------------------------------
\section{Introduction}\label{sintro}
%----------------------------------------------------------------------

Consider a large set of $n$ units, and for each unit $i$, let $Y_i$ denote its outcome and $D_i$ its treatment assignment. We study settings with interference in which outcomes may depend on the entire treatment assignment vector $\bm{D} = (D_i)_{i=1}^n \in \{0,1\}^n$. Because the causal effect of $\bm{D}$ on $Y_i$ is difficult to convey, let alone identify, a common strategy is to regress $Y_i$ on a substantially lower-dimensional vector $T_i$ that parsimoniously summarizes $\bm{D}$, for example the number of treated neighbors. We consider nonparametric analogs of these regression estimands, which take the form
\begin{equation*}
  \tau(t,t') = \frac{1}{n} \sum_{i=1}^n \big( \E[Y_i \mid T_i=t, \C_i] - \E[Y_i \mid T_i=t', \C_i] \big).\footnote{As is convention, throughout the paper we use the notation $\E[Y_i \mid T_i=t, \C_i]$ to mean $m_t(\C_i)$ for $m_t(c) = \E[Y_i \mid T_i=t, \C_i=c]$.}
\end{equation*}

\noindent If $T_i$ counts $i$'s treated neighbors, then $\tau(t,t')$ compares the average outcomes of units with different numbers of treated neighbors, controlling for $\C_i$. The literature refers to $T_i$ as an {\em effective treatment} or {\em exposure mapping} \citep{manski2013identification,aronow2017estimating}. We refer to $\tau(t,t')$ as an {\em exposure contrast} and study when and in what sense it has a causal interpretation.

An example of this empirical strategy is \cite{lu2019place} who study the impact of Chinese special economic zones (SEZs) on village-level outcomes. In their setting, $D_i=1$ if village $i$ is situated in an SEZ. Because there may be spillovers from neighboring SEZs, the authors regress outcomes on an exposure mapping that includes $D_i$ and an indicator for having a treated neighbor, that is, a village in $i$'s county that lies in an SEZ. \cite{baird2018optimal}, \cite{cai2015social}, and \cite{miguel2004worms} instead measure spillovers using the number or share of treated peers within a neighborhood or cluster.\footnote{\cite{donaldson2016railroads}, \cite{kline2014local}, and \cite{zheng2017birth} employ similar strategies.} In all cases, the coefficient on own treatment $D_i$ is intended to capture a direct effect while the coefficient on the statistic involving neighbors' treatments is intended to capture a spillover effect. The question is whether these interpretations are warranted.

Regressions of this sort are likely intended as devices for producing summary measures of spillover effects (i.e.\ exposure contrasts) rather than as structural models of interference. Yet the causal literature predominantly treats the exposure mapping as structural in the sense that it entirely summarizes the effect of $\bm{D}$ on $Y_i$. For most exposure mappings used in the literature, this is incompatible with endogenous peer effects, a leading explanation for interference in many social and economic contexts \citep{jackson2022inequality,sacerdote2011peer}, since outcomes depend on the entirety of $\bm{D}$ under the reduced form of a simultaneous-equations model. It stands in contrast to a large literature on social interactions that specifies structural models with endogenous peer effects.\footnote{E.g.\ \cite{blume2015linear}, \cite{bramoulle2009identification}, \cite{lazzati2015treatment}, \cite{lewbel2023social}, and \cite{manski1993identification}.} These models have richer microfoundations, but point identification typically relies on parametric assumptions. To bridge the two approaches, this paper studies the causal interpretation of exposure contrasts in a nonparametric setting where exposures are not structural. 

We demonstrate that, even when assignment is unconfounded, an exposure contrast can have the opposite sign of the unit-level effects. A sign reversal is possible if (i) interference is more complex than what the exposure mapping dictates and (ii) treatments are correlated across units. We then provide interpretable nonparametric restrictions on either (i) or (ii) that, together with unconfoundedness, ensure that common exposure contrasts can be represented as convex averages of the unit-level effects. 

Our first identification result pertains to ``count'' exposures such as the number of treated neighbors. We provide conditions under which there exist partially ordered treatment assignment vectors $\bm{D}_{i,t}^* \stackrel{a.s.}\geq \bm{D}_{i,t'}^*$ for all $i$ such that the exposure contrast can be written as an average of unit-level effects $n^{-1} \sum_{i=1}^n \E[Y_i(\bm{D}_{i,t}^*) - Y_i(\bm{D}_{i,t'}^*) \mid \C_i]$ (see \autoref{scount}). The result imposes no restrictions on interference but requires treatment assignments to be conditionally independent. We show that the latter holds when selection into treatment is governed by a game of incomplete information.

Our second result considers estimands for cluster-randomized trials which compare units in clusters assigned to distinct saturation levels. For example the ``overall effect'' compares mortality rates between clusters with different vaccination rates. These have causal interpretations under the restrictive ``stratified interference'' assumption that the share of treated peers entirely mediates interference, but their interpretations more generally have not been studied. We derive convex average representations without imposing any restrictions on interference within or across clusters (see \autoref{sCRT}).

The previous results utilize restrictions on (ii). We also provide results that leave the assignment mechanism unrestricted and instead impose assumptions on interference. One restriction requires spillovers beyond the $K$-neighborhood to be of smaller order than those within the $K$-neighborhood. Another supposes that spillovers decay with network distance (see \autoref{stselect}). Both are weaker than the assumption of structural exposure contrasts and allow for endogenous peer effects.

A convex average representation provides a sense in which an exposure contrast can be considered causal, but whether it is ``policy relevant'' is a different matter \citep{auerbach2024discussion}. Under the stable unit treatment value assumption (SUTVA), common policy effects are convex averages of unit-level effects with specific weights \citep{heckman2007econometric}. Identifying analogous policy effects under interference would presumably require conditions that eliminate sign reversals, which this paper provides.

\bigskip
\noindent {\bf Related Literature.} Compared to the peer effects literature, the unit-level effects we study are reduced-form in that they do not distinguish between endogenous and exogenous peer effects \citep{manski1993identification}. The upshot is that identification is possible without imposing parametric structure, in the spirit of \cite{manski2013identification}. Several of our results allow for peer effects in outcomes and selection into treatment. \cite{balat2023multiple} also consider strategic interactions in selection and derive bounds on the average treatment effect.

A large literature studies the causal interpretations of various regression estimands under SUTVA when treatment effects are heterogeneous \citep[e.g.][]{blandhol2022tsls,bugni2023decomposition,de2020two,goldsmith2024contamination,small2017instrumental}. Sign reversals can occur due to the use of linear regression, and reversals can be avoided by using estimators directly targeting nonparametric estimands. This is not the case in our setting. We study a nonparametric estimand, and reversals occur not due to heterogeneity but rather the combination of interference and correlated treatments across units.

\cite{savje2024causal} observes that exposure mappings serve two distinct roles in the literature: ``to define the effect of interest and to impose assumptions on\ldots interference.'' He argues that they should only be used for the former purpose. He refers to $\tau(t,t')$ as the ``expected exposure effect'' and to the special case of $T_i=D_i$ as the ``average distributional shift effect'' (his \S S2), implying that $\tau(t,t')$ has a causal interpretation when treatment is randomized. To the contrary, we show that, even under randomized assignment, $\tau(t,t')$ can exhibit unpalatable sign reversals that are only avoided under additional restrictions.\footnote{In \autoref{ssignp} of the appendix, we discuss the sign preservation criterion proposed by \cite{savje2024rejoinder} and how it relates to our results.}

Our identification results complement work on estimation and inference for exposure contrasts when exposure mappings are not structural. \cite{leung2022causal} and \cite{leung2024graph} study large-sample inference on $\tau(t,t')$ under asymptotics sending $n\rightarrow\infty$. \cite{savje2024causal} states high-level conditions for consistent estimation. None formally study the causal interpretation of $\tau(t,t')$.\footnote{The first working paper version of \cite{leung2022causal} provides a limited formal discussion of the causal interpretation under Bernoulli-randomized designs \citep[][\S A.1]{leung2019causal}. An earlier draft of \cite{leung2024graph} included some of the identification results in \autoref{sKnbhd}.}

\bigskip
\noindent {\bf Outline.} We describe the basic setup in the next section and subsequently organize results by the classes of exposure mappings to which they pertain. Section \ref{scount} considers exposures involving counts of treated units and provides a motivating sign reversal example. In \autoref{sCRT}, we turn to exposure contrasts common in the literature on cluster-randomized trials. In \autoref{sKnbhd}, we study $K$-neighborhood exposure mappings, which summarize the treatment configuration within a local network neighborhood. Section \ref{sconclude} concludes. Proofs of these results can be found in \autoref{sproofs}.

%----------------------------------------------------------------------
\section{Setup}\label{sset}
%----------------------------------------------------------------------

Recall that $\bm{D} = (D_i)_{i=1}^n \in \{0,1\}^n$ is the observed assignment vector. We refer to its distribution as the {\em assignment mechanism}.  For all $i \in \N_n = \{1,\ldots,n\}$, let $Y_i(\cdot)$ be a (potentially) random mapping from $\{0,1\}^n$ to $\R$. We interpret $Y_i(\bm{d})$ as the potential outcome of unit $i$ under the counterfactual that the treatment assignment vector is $\bm{d} = (d_i)_{i=1}^n \in \{0,1\}^n$ so that the observed outcome $Y_i$ equals $Y_i(\bm{D})$. Interference arises because potential outcomes may depend not only on own assignment $D_i$ but also on the entire assignment vector. 

Let $\C_i$ denote an array of control variables for unit $i$, the choice of which we discuss in \autoref{sCI}. We assume treatment assignments are unconfounded in the following sense.

\begin{myassump}{UC}\label{aunc} 
  $Y_i(\cdot) \indep \bm{D} \mid \C_i$ for all $i\in\N_n$.
\end{myassump}

We primarily consider exposure mappings that are deterministic functions of the assignment vector: 
\begin{equation*}
  T_i = f(i,\bm{D}) \quad\text{for some}\quad f\colon \N_n \times \{0,1\}^n \rightarrow \R^{d_t}.
\end{equation*}

\noindent Most of the literature assumes the exposure mapping is {\em structural} in the following sense \citep[e.g.][]{aronow2017estimating,forastiere2021identification,ogburn2024causal}.

\begin{definition}\label{dstr}
  An exposure mapping $f$ is {\em structural} if $Y_i(\bm{d}) = Y_i(\bm{d}')$ for all $i$ and $\bm{d},\bm{d}' \in \{0,1\}^n$ such that $f(i,\bm{d}) = f(i,\bm{d}')$.
\end{definition}

\noindent In this case, we can rewrite potential outcomes as $Y_i(f(i,\bm{d}))$, so under \autoref{aunc}, $\tau(t,t')$ reduces to $n^{-1} \sum_{i=1}^n \E[Y_i(t) - Y_i(t') \mid \C_i]$, which has a transparent causal interpretation. The primary purpose of this assumption is dimension-reduction by specifying $f$ with dimension much smaller than $n$, typically 1 or 2 in practice. This is necessary for $\tau(t,t')$ to be estimable without substantial additional restrictions. 

However, when researchers articulate why spillovers exist in a given setting, it is frequently a story of outcomes affecting outcomes, of peer decisions or behavior influencing those of the ego. In other words, it is a story of endogenous peer effects. In this case, as discussed in the introduction, \autoref{dstr} can only hold for the trivial exposure mapping $f(i,\bm{d}) = \bm{d}$ which has no dimension-reduction benefit. In this paper, we take as given the researcher's choice of a low-dimensional summary $f$ and study under what conditions $\tau(t,t')$ retains a causal interpretation when $f$ is not structural. We will consider weaker assumptions allowing for complex forms of interference such as endogenous peer effects.

Throughout the paper we maintain the {\em overlap condition} that the conditional distribution $\bm{D} \mid T_i=s, \C_i=c$ exists for all $c$ in the support of $\C_i$, $i\in\N_n$, and $s \in \{t,t'\}$. For instance if $T_i$ is the number of $i$'s treated neighbors, $t=2$, and $t'=1$, then overlap implies that the exposure contrast only averages over units $i$ with at least 2 neighbors.\footnote{We leave this implicit in the notation since the neighborhood structure is typically treated as fixed or conditioned upon.}

\begin{remark}
  This paper is concerned with identification, and as such, we treat $\tau(t,t')$ as known to the econometrician. \cite{leung2024graph} studies doubly robust estimation of $\tau(t,t')$ under network interference and asymptotics sending the number of units $n$ to infinity. \cite{leung2025cluster} studies cluster-randomized trials with spatial interference under the same asymptotics. Both papers require additional conditions for weak dependence that are not necessary for most of our results. The main condition is that interference decays sufficiently quickly with distance (see \autoref{aani} in \autoref{stselect}). This is substantially weaker than assuming a structural exposure mapping and allows for endogenous peer effects \citep{leung2022causal}.
\end{remark}

%----------------------------------------------------------------------
\section{Count Exposures}\label{scount}
%----------------------------------------------------------------------

This section considers the following class of exposure mappings. For $I_1, I_2 \subseteq \R$, we write $I_1 \leq I_2$ if $a \leq b$ for all $a \in I_1$ and $b \in I_2$.

\begin{myassump}{CT}\label{aCT}
  For any $i \in \N_n$, there exist mutually exclusive sets $B_{i1}, \ldots, B_{id_t} \subseteq \N_n$ and possibly unbounded intervals satisfying $I_{i1}^{(t)} \geq I_{i1}^{(t')}, \ldots, I_{id_t}^{(t)} \geq I_{id_t}^{(t')}$ such that for all $s \in \{t,t'\}$,
  \begin{equation*}
    \ind\{f(i,\bm{d}) = s\} = \prod_{r=1}^{d_t} \ind\bigg\{ \sum_{j \in B_{ir}} d_j \in I_{ir}^{(s)} \bigg\}.
  \end{equation*}
\end{myassump}

When the intervals are singleton sets, this says $\ind\{f(i,\bm{d}) = s\}$ is an indicator for whether a vector of treated unit counts takes on a particular value. 

\begin{example}\label{enc}
  Call $f(i,\bm{d}) = (d_i, \sum_{j=1}^n A_{ij} d_j)$ the {\em treated neighbor count}, where $A_{ij}$ is a non-random indicator for whether units $i$ and $j$ are neighbors. Neighbors could be social contacts, units in the same ``cluster,'' units within a certain geographic distance band, etc. Let $t = (0, t_2)$ and $t' = (0, t_2')$ with $t_2 \geq t_2'$, so $\tau(t,t')$ compares untreated units with $t_2$ and $t_2'$ treated neighbors. This satisfies \autoref{aCT} with $d_t=2$, $B_{i1}=\{i\}$, $B_{i2}=\{j\in\N_n\colon A_{ij}=1\}$, $I_{i1}^{(s)}=\{0\}$ for $s \in \{t,t'\}$, $I_{i2}^{(t)} = \{t_2\}$, and $I_{i2}^{(t')}=\{t_2'\}$.
\end{example}

\begin{example}
  In place of the count $\sum_{j=1}^n A_{ij} d_j$, some references use the indicator $\ind\{\sum_{j=1}^n A_{ij} d_j > 0\}$ \citep[e.g.][]{lu2019place}. Then if $t = (1, t_2)$ and $t' = (1, t_2')$, $t_2=1$, and $t_2'=0$, $\tau(t,t')$ compares treated units with some and no treated neighbors. This satisfies \autoref{aCT} with $d_t=2$, $B_{i1}=\{i\}$, $B_{i2}=\{j\in\N_n\colon A_{ij}=1\}$, $I_{i1}^{(s)}=\{1\}$ for $s \in \{t,t'\}$, $I_{i2}^{(t)} = (0,\infty)$, and $I_{i2}^{(t')} = \{0\}$.
\end{example}

The exposure contrasts in these examples are intended to capture a spillover effect from varying the number of treated neighbors. The question is what justifies such an interpretation. The next subsection shows that this contrast generally does not preserve the sign of unit-level spillover effects even under unconfoundedness. In \autoref{srep}, we provide a restriction on the assignment mechanism under which $\tau(t,t')$ is a convex average of the unit-level effects and hence avoids unpalatable sign reversals. In the last subsection, we provide examples of assignment mechanisms that satisfy the restriction.

%---------------------------------------------
\subsection{Sign Reversal}\label{srev}
%---------------------------------------------

Consider a setting with no control variables $\C_i$ and $n/4$ identical clusters, each with 4 units (``neighbors'') labeled 1--4. Within a cluster, the potential outcome of a unit $i \in \{1,\ldots,4\}$ is denoted by $Y_i(d_1,d_2,d_3,d_4)$ where $d_j$ denotes the treatment of the $j$th neighbor. We leave the cluster index implicit on account of clusters being identical. 

Consider the treated neighbor count from \autoref{enc}. Let $t=(1,2)$ and $t'=(1,1)$, so $\tau(t,t')$ compares average outcomes of treated units with either 2 or 1 treated neighbors. The unit-level effects of interest then include comparisons such as
\begin{equation*}
  Y_1(1,1,1,0) - Y_1(1,1,0,0). 
\end{equation*}

\noindent This is the spillover effect for a treated unit from having one additional neighbor treated relative to a baseline of 1 treated neighbor. Contrast this with 
\begin{equation}
  Y_1(1,1,1,0) - Y_1(1,0,0,1) \label{forbidden}
\end{equation}

\noindent which involves moving neighbor 4 out of treatment and the other neighbors into treatment. Our position is that the exposure contrast is intended to capture the effect of moving one additional unit into treatment, not simultaneously switching neighbors in and out of treatment. The distinguishing feature of the first comparison is that the assignment vectors are partially ordered, unlike those of the second. Thus in general when $t \geq t'$, the unit-level comparisons of interest are effects of the form
\begin{equation}
  Y_i(\bm{d}) - Y_i(\bm{d}') \quad\text{s.t.}\quad \bm{d} \geq \bm{d}', \quad f(i,\bm{d})=t, \quad f(i,\bm{d}')=t'. \label{target}
\end{equation}

We next demonstrate that the sign of $\tau(t,t')$ can be entirely inconsistent with the signs of \eqref{target} even under a randomized control trial. Suppose potential outcomes are given by
\hfill
\begin{center}
  \begin{tabular}{lll}
    $Y_1(1,1,1,0) = 1.5$ & & $Y_1(1,1,0,0) = 0$ \\
    $Y_1(1,1,0,1) = 2.5$ & & $Y_1(1,0,1,0) = 1$ \\
    $Y_1(1,0,1,1) = 3$ & & $Y_1(1,0,0,1) = 2$
  \end{tabular}
\end{center}

\hfill

\noindent and $Y_i(\bm{d})=0$ for all other $\bm{d}$ and $i\neq 1$. Observe that all unit-level effects of the form \eqref{target} are non-negative and, for unit 1, strictly positive. Consider the cluster-randomized trial that assigns treatments independently across clusters, such that the distribution of within-cluster treatments only places positive probability on the vectors $(1,1,1,0)$ and $(1,0,0,1)$. Since clusters are identical with four units each, 
\begin{equation*}
  \tau(t,t') = \frac{1}{4}\big(\E[Y_1 \mid T_1=(1,2)] - \E[Y_1 \mid T_1=(1,1)]\big) = \frac{1}{4}(1.5 - 2) < 0.
\end{equation*}

The sign reversal occurs because the exposure mapping is not structural, and treatment assignments are correlated across units. If the exposure mapping were structural, which is a restriction on interference, then only the number of treated neighbors, not their identities, would matter for potential outcomes, and no sign reversal would occur, per the discussion following \autoref{dstr}. If treatments were i.i.d., which is a restriction on the assignment mechanism, then one can calculate that $\tau(t,t') > 0$.\footnote{Let $\bm{D}_{(j)}$ be the treatment subvector of any cluster $j$. If treatments are i.i.d.\ across units, then by symmetry $\prob(\bm{D}_{(j)}=\bm{d} \mid T_1=(1,2)) = 1/3$ for all $\bm{d}$ such that $f(1,\bm{d})=(1,2)$, and $\prob(\bm{D}_{(j)}=\bm{d}' \mid T_1=(1,1)) = 1/3$ for all $\bm{d}'$ such that $f(1,\bm{d}')=(1,1)$, so $\tau(t,t') = (1.5/3+2.5/3+1 - 0 - 1/3 - 2/3)/4 > 0$.}

The remainder of the paper considers weaker restrictions on interference and the assignment mechanism under which $\tau(t,t')$ avoids sign reversals. The first three theorems leave interference entirely unrestricted, while the last two leave the assignment mechanism unrestricted. All results allow for endogenous peer effects, unlike the assumption of structural exposures.

%---------------------------------------------
\subsection{Representation Result}\label{srep}
%---------------------------------------------

Our first result exploits the following restriction on the assignment mechanism.

\begin{myassump}{CI}\label{aCI0}
  The components of $\bm{D}$ are independent conditional on $\C_i$, for any $i \in \N_n$.
\end{myassump}

\noindent Notice this does not require treatments to be identically distributed, which will be important for applications to observational data. We state the main result and defer further discussion of the assumption to the next subsection. Let $p_{i,t}(\cdot \mid c)$ denote the conditional probability mass function (PMF) of $\bm{D}$ given $T_i=t, \C_i=c$.

\begin{theorem}\label{tcount}
  Under Assumptions \ref{aunc}, \ref{aCT}, and \ref{aCI0}, for all $i \in \N_n$ there exists a monotone coupling $\bm{D}_{i,t}^* \stackrel{a.s.}\geq \bm{D}_{i,t'}^*$ with $\bm{D}_{i,s}^* \sim p_{i,s}(\cdot \mid \C_i)$ for all $s \in \{t,t'\}$ such that 
  \begin{equation*}
    \tau(t,t') = \frac{1}{n} \sum_{i=1}^n \E\big[ Y_i(\bm{D}_{i,t}^*) - Y_i(\bm{D}_{i,t'}^*) \mid \C_i \big]. 
  \end{equation*}
\end{theorem}

\noindent Since $\bm{D}_{i,s}^* \sim p_{i,s}(\cdot \mid \C_i)$, we have $f(i,\bm{D}_{i,s}^*)=s$ a.s. Thus the result shows that exposure contrasts can be represented as convex averages of the unit-level effects in \eqref{target}. The weights in the average are determined by the conditional distributions of assignment vectors $p_{i,s}(\cdot \mid \C_i)$ for $s \in \{t,t'\}$.

By definition, $\tau(t,t')$ is a difference of two convex averages, and under unconfoundedness, this can always be represented as a convex average of differences if we include all possible unit-level comparisons of the form $Y_i(\bm{d})-Y_i(\bm{d}')$ in the average (see the proof of \autoref{pGSP}). This means including comparisons discussed in \autoref{srev} that are undesirable because they involve simultaneously moving units into and out of treatment. \autoref{tcount} shows that these comparisons are properly excluded under \autoref{aCI0}.

\begin{remark}
  In \autoref{enc}, it appears natural to interpret $\tau(t,t')$ as a spillover effect of increasing the number of treated neighbors from $t_2'$ to $t_2$. \autoref{tcount} guarantees such an interpretation. Let $\N(i,1) = \{j \in \N_n\colon A_{ij}=1\}$, and denote by $\bm{D}_B$ and $\bm{D}_{-B}$ the subvectors of $\bm{D}$ on $B$ and $\N_n\backslash B$, respectively. Let $p^1_{i,s}(\cdot \mid c)$ be the PMF of $\bm{D}_{\N(i,1)}$ given $T_i=s, \C_i=c$. Since treatments are conditionally independent, the theorem implies the existence of a monotone coupling $\bm{D}_{\N(i,1),t}^* \stackrel{a.s.}\geq \bm{D}_{\N(i,1),t'}^*$ independent of $\bm{D}_{-\N(i,1)}$ with $\bm{D}_{\N(i,1),s}^* \sim p^1_{i,s}(\cdot \mid \C_i)$ for all $s \in \{t,t'\}$ such that 
  \begin{equation*}
    \tau(t,t') = \frac{1}{n} \sum_{i=1}^n \E\big[ Y_i(\bm{D}_{\N(i,1),t}^*, \bm{D}_{-\N(i,1)}) - Y_i(\bm{D}_{\N(i,1),t'}^*, \bm{D}_{-\N(i,1)}) \mid \C_i \big]. 
  \end{equation*}

  \noindent Given $t=(0,t_2)$ and $t'=(0,t_2')$, $\bm{D}_{\N(i,1),t}^*$ is a vector with $i$ untreated and $t_2$ treated neighbors, while $\bm{D}_{\N(i,1),t'}^*$ is the vector with $i$ untreated and $t_2'$ treated neighbors.
\end{remark}

\begin{remark}
  In the difference-in-differences literature, the estimands of two-way fixed effects regressions can exhibit sign reversals because they contain both ``clean'' and ``forbidden'' unit-level comparisons \citep{roth2023whats}. The solution in this literature is to define new nonparametric estimands that only contain clean comparisons. The exposure contrast is a nonparametric estimand, which \autoref{srev} shows generally contains both ``clean'' \eqref{target} and ``forbidden'' \eqref{forbidden} comparisons. Rather than defining a new estimand, \autoref{tcount} provides conditions under which ``forbidden'' comparisons are excluded from the exposure contrast.
\end{remark}

\begin{remark}
  The example in \autoref{srev} shows that \autoref{tcount} does not extend to arbitrary dependent treatment assignments. We next show it does not extend either to arbitrary nondecreasing or even additive exposure mappings. Suppose $n=3$, treatments are independent across units, and $f(i,\bm{d}) = 3d_i + 2 \sum_{j\neq i} d_j$, a weighted sum of treatments. Consider potential outcomes
  \begin{center}
    \begin{tabular}{lll}
      $Y_1(1,1,1) = 2$ & $Y_1(1,1,0) = 1.5$ & $Y_1(1,0,1) = 1$ \\
      $Y_1(0,1,1) = 0$ & $Y_1(1,0,0) = 1$ & 
    \end{tabular}
  \end{center}

  \noindent and $Y_i(\bm{d})=0$ in all other cases. Then all unit-level treatment and spillover effects (e.g.\ $Y_1(1,d_2,d_3)-Y_1(0,d_2,d_3)$ and $Y_1(d_1,1,d_3)-Y_1(d_1,0,d_3)$) are non-negative, yet $\tau(4,3) = (\E[Y_1 \mid T_1=4] - \E[Y_1 \mid T_1=3])/3 = (Y_1(0,1,1) - Y_1(1,0,0))/3 = -1/3$.
\end{remark}

\begin{remark}
  The proof of \autoref{tcount} relies specifically on count statistics due to the fact that sums $T$ of independent Bernoulli random variables $\bm{D} = (D_1, \ldots, D_n)$ have log-concave PMFs (\autoref{dlc}). Log-concavity enables application of a useful monotonicity result due to \cite{efron1965increasing} (\autoref{lefron}) to show that the distribution of $\bm{D} \mid T = t$ stochastically dominates that of $\bm{D} \mid T = t'$.
\end{remark}

%---------------------------------------------
\subsection{\autoref{aCI0}}\label{sCI}
%---------------------------------------------

We next discuss examples of assignment mechanisms satisfying \autoref{aCI0}.

\bigskip

\noindent {\bf Experimental Data.} There is a growing literature on experimental design under network interference. Network targeting experiments may randomize treatments among the subset of nodes with a particular local network configuration, sometimes referred to as ``seeds'' or ``injection points'' \citep[e.g.][]{beaman2021can,kim2015social}. Treatments are then functions of the network $\bm{A}$ and possibly unit-level covariates $\bm{X}$, but remain independent conditional on $(\bm{X},\bm{A})$. Then \autoref{aCI0} holds with $\C_i = (\bm{X},\bm{A})$ for all $i$. We discuss controls of this sort further below.

Many proposed designs induce correlation in assignments beyond stratification, for example the balancing designs of \cite{basse2018model}, the independent-set design of \cite{karwa2018systematic}, and the quasi-coloring design of \cite{jagadeesan2020designs}. These papers all assume particular structural exposure mappings. \autoref{tcount} provides a reason to prefer conditionally independent designs, namely to ensure that the causal interpretations of their estimands are robust to complex forms of interference. In \autoref{stselect}, we state results that leave the assignment mechanism unrestricted and hence allow for correlated designs.

\bigskip

\noindent {\bf Observational Data.} \cite{leung2024graph} propose a nonparametric model of network interference that allows for strategic interactions in both the outcome stage and selection stage. Let
\begin{equation}
  Y_i = g_n(i,\bm{D},\bm{X},\bm{A},\bm{\varepsilon}) \quad\text{and}\quad D_i = h_n(i,\bm{X},\bm{A},\bm{\nu}) \label{YD}
\end{equation}

\noindent for all $i\in\N_n$, where $\bm{A}$ is the network (formally an $n\times n$ matrix), $\bm{X} = (X_i)_{i=1}^n$ an array of unit-level observables, $\bm{\varepsilon} = (\varepsilon_i)_{i=1}^n$ an array of outcome unobservables, $\bm{\nu} = (\nu_i)_{i=1}^n$ an array of selection unobservables, and $\{(g_n,h_n)\}_{n\in\mathbb{N}}$ a sequence of function pairs such that each $g_n(\cdot)$ has range $\R$ and $h_n(\cdot)$ has range $\{0,1\}$. The timing of the model is that nature draws $(\bm{A},\bm{X},\bm{\varepsilon},\bm{\nu})$; units select into treatment according to a simultaneous-equations model with reduced form $h_n(\cdot)$; and outcomes are realized according to a simultaneous-equations model with reduced form $g_n(\cdot)$. 

\begin{example}\label{ebg2}
  Suppose selection into treatment is determined by a game of incomplete information in which units take up treatment to maximize expected utility
  \begin{equation}
    D_i = \bm{1}\big\{ \E_i[U_i(\bm{D}_{-i}, \bm{X}, \bm{A}, \bm{\nu}) \mid \bm{X}, \bm{A}, \nu_i] > 0 \big\} \label{minc}
  \end{equation}

  \noindent where $(\bm{X},\bm{A},\nu_i)$ is the information set of unit $i$ and $\E_i[\cdot]$ is the expectation taken with respect to $i$'s beliefs \citep{bajari2010estimating,xu2018social}. Under the usual assumption that equilibrium selection only depends on public information $(\bm{X},\bm{A})$, the selection model can be represented as
  \begin{equation}
    D_i = h_n(i,\bm{X},\bm{A},\nu_i).
    \label{incinf}
  \end{equation}
\end{example}

Under model \eqref{YD}, potential outcomes are given by $Y_i(\bm{d}) = g_n(i,\bm{d},\bm{X},\bm{A},\bm{\varepsilon})$. Since these and treatments depend on the entirety of $(\bm{X},\bm{A})$, to account for high-dimensional network confounding, we take 
\begin{equation}
  \C_i = (\bm{X},\bm{A}) \quad\text{for all } i\in\N_n. \label{confounders}
\end{equation}

\noindent \cite{leung2024graph} provide conditions under which doubly-robust estimation of exposure contrasts is feasible with controls \eqref{confounders}.

\begin{proposition}\label{pinc}
  Suppose the assignment mechanism is governed by model \eqref{incinf}, and controls are given by \eqref{confounders}. If $\{\nu_j\}_{j=1}^n$ is independently distributed conditional on $(\bm{X},\bm{A})$, then so is $\{D_i\}_{i=1}^n$.
\end{proposition}

\noindent The proof is straightforward and omitted. The empirical games literature studying estimation of \eqref{minc} under large-market asymptotics typically assumes private information is i.i.d.\ and independent of public information \citep[e.g.][]{lin2021selection,lin2017estimation,xu2018social}. This implies the conditional independence restriction in the theorem. 

By \autoref{tcount} and \autoref{pinc}, $\tau(t,t')$ obtains a convex average representation without any restrictions on interference or the magnitude of peer effects in selection, provided we model selection (nonparametrically) as a certain game of incomplete information. 

%----------------------------------------------------------------------
\section{Saturation Exposures}\label{sCRT}
%----------------------------------------------------------------------

We next turn to exposure contrasts common in the cluster-randomized trials (CRTs) literature. We suppose treatments are assigned according to a standard randomized saturation design.

\begin{myassump}{CRT}\label{aCRT}
  For all $i\in\N_n$, $Y_i(\cdot) \indep \bm{D}$. Let the clusters $\{C_j\}_{j=1}^m$ be a partition of $\N_n$ and the saturation levels $\{\tilde{S}_j\}_{j=1}^m$ be i.i.d.\ draws from a distribution supported on $\mathcal{P} = \{p_k\}_{k=1}^q \subseteq [0,1]$. For each $j$, $\{D_i\}_{i \in C_j} \stackrel{iid}\sim \text{Bernoulli}(p)$ conditional on $\tilde{S}_j = p$. 
\end{myassump}

\noindent That is, clusters are randomly assigned to saturation levels, and units within a cluster are assigned to treatment with probability equal to the saturation level. Let $S_i$ be the saturation level assigned to unit $i$'s cluster, so that $S_i = \tilde{S}_j$ if $i \in C_j$. 

In this section, we depart from the setup of \autoref{sset} and define the exposure mapping as the tuple
\begin{equation*}
  T_i = (D_i, S_i).
\end{equation*}

\noindent Most of the CRT literature focuses on the following four exposure contrasts $\tau(t,t')$ \citep{hayes2017cluster}.
\begin{enumerate}
  \item The ``direct effect'' sets $t=(1,p)$ and $t'=(0,p)$ for any $p \in \mathcal{P}$, meaning it conditions on the saturation level but varies own treatment assignment. Under Assumptions \ref{aunc} and \ref{aCRT}, this has a clear causal interpretation due to Bernoulli-randomization within cluster. This is not the case for the remaining estimands.

  \item The ``indirect effect'' sets $t=(d,p)$ and $t'=(d,p')$ for any $d \in \{0,1\}$ and $p,p' \in \mathcal{P}$ with $p\geq p'$, thus varying the saturation level while conditioning on the treatment. 

  \item The ``total effect'' is the sum of the direct and indirect effects, which corresponds to setting $t = (1,p)$ and $t'= (0,p')$. 

  \item The ``overall effect,'' for lack of better notation, sets $t = (\emptyset,p)$ and $t' = (\emptyset,p')$ where we define the event $\{D_i = \emptyset\} \equiv \{D_i \in \{0,1\}\}$. In other words, the contrast varies the saturation level without conditioning on treatment assignment $D_i$.\footnote{The ``group average effects'' of \cite{hudgens2008toward} are conceptually similar to these definitions. The main distinction is that our definition of the exposure contrast equally weights units whereas theirs equally weights clusters.}
\end{enumerate}

In all cases, $\tau(t,t')$ is only a statistical comparison and potentially subject to sign reversals of the sort in \autoref{srev}. A common assumption in the CRT literature is {\em stratified interference}, which just says that the exposure mapping is structural \citep[e.g.][]{basse2018analyzing,hudgens2008toward,vazquez2023identification}. Then we can rewrite $Y_i(\bm{D})$ as $Y_i(T_i)$, and the four ``effects'' have transparent causal interpretations. However this is restrictive because it presumes units are exchangeable within cluster. Units may instead respond differently to others depending on characteristics or social connections.

Perhaps for this reason, some papers do not maintain stratified interference \citep[e.g.][]{hudgens2008toward,lee2024efficient,tchetgen2012causal}, but the causal meaning of the estimands in this case has not been studied in the literature. Furthermore, virtually all references assume {\em partial interference}, meaning there is no interference across clusters, but cross-cluster interference is often a feature of CRTs for infectious diseases and large-scale social experiments \citep{egger2022general,leung2025cluster}. 

The next result shows that, for standard designs satisfying \autoref{aCRT}, no restrictions on interference are required to ensure that the estimands can be represented as convex averages of unit-level effects. Let $C_{(i)}$ denote the cluster containing unit $i$, and for any $\bm{d} \in \{0,1\}^n$, let $\bm{d}_{(i)} = (d_j)_{j \in C_{(i)}}$ and $\bm{d}_{(-i)} = (d_j)_{j \in \N_n\backslash C_{(i)}}$. Finally let $p_{(i),s}(\cdot)$ denote the conditional distribution of $\bm{D}_{(i)} \mid T_i=s$ for $s \in \{t,t'\}$.

\begin{theorem}\label{tCRT}
  Under \autoref{aCRT}, if $\tau(t,t')$ is the indirect, total, or overall effect with $p\geq p'$, then for all $i\in\N_n$ there exists a monotone coupling $\bm{D}_{(i),t}^* \stackrel{a.s.}\geq \bm{D}_{(i),t'}^*$ independent of $\bm{D}_{(-i)}$ with $\bm{D}_{(i),s}^* \sim p_{(i),s}(\cdot)$ for all $s \in \{t,t'\}$ such that 
  \begin{equation*}
    \tau(t,t') = \frac{1}{n} \sum_{i=1}^n \E\big[ Y_i(\bm{D}_{(i),t}^*, \bm{D}_{(-i)}) - Y_i(\bm{D}_{(i),t'}^*, \bm{D}_{(-i)}) \big]. 
  \end{equation*}
\end{theorem}

\noindent Because $p\geq p'$, an increase in the exposure $T_i$ from $t'$ to $t$ means an increase in the proportion treated, resulting in stochastically larger assignment vectors $\bm{d}_{(i)}$ in $i$'s cluster. Because treatments are independent across clusters, this induces no change in $\bm{d}_{(-i)}$. The unit-level effects of interest therefore take the form $Y_i(\bm{d}_{(i)}, \bm{d}_{(-i)}) - Y_i(\bm{d}_{(i)}', \bm{d}_{(-i)})$ with $\bm{d}_{(i)} \geq \bm{d}_{(i)}'$, which are exactly those in the convex average. 

%----------------------------------------------------------------------
\section{$K$-Neighborhood Exposures}\label{sKnbhd}
%----------------------------------------------------------------------

Suppose units are connected through a network $\bm{A}$, represented as an $n\times n$ binary matrix with $ij$th entry $A_{ij}$ which we treat as non-random or conditioned upon. Let $\N(i,K)$ denote unit $i$'s {\em $K$-neighborhood}, the subset of units at most path distance $K$ from $i$ in $\bm{A}$.\footnote{The path distance between two distinct units is the length of the shortest path between them if a path exists and infinite if not. The path distance between a unit and itself is zero, so $\N(i,0) = \{i\}$.} For any $\bm{d} \in \{0,1\}^n$, let $\bm{d}_{\mathcal{N}(i,K)} = (d_j\colon j \in \N(i,K))$ and $\bm{d}_{-\N(i,K)} = (d_j\colon j \in \N_n\backslash\N(i,K))$. It will often be convenient to partition $\bm{d}$ as $(\bm{d}_{\N(i,K)}, \bm{d}_{-\N(i,K)})$ and write $Y_i(\bm{d}_{\N(i,K)}, \bm{d}_{-\N(i,K)}) \equiv Y_i(\bm{d})$. 

This section considers the setup of \autoref{sset}, with the additional restriction that $f$ is a {\em $K$-neighborhood} exposure mapping in that it only depends on treatments assigned to the ego's $K$-neighborhood. Formally, $f(i,\bm{d}) = f(i,\bm{d}')$ for all $i$ and $\bm{d},\bm{d}' \in \{0,1\}^n$ such that $\bm{d}_{\mathcal{N}(i,K)} = \bm{d}_{\mathcal{N}(i,K)}'$. Abusing notation, we may abbreviate
\begin{equation*}
  f(\bm{d}_{\N(i,K)}) \equiv f(i,\bm{d}).
\end{equation*}

\noindent The treated neighbor count in \autoref{enc} satisfies this restriction with $K=1$. 

If $K$ is chosen large enough to encompass the entire network, this imposes no restrictions. However, the spirit of exposure mappings is to choose $K$ smaller than the typical distance between units to parsimoniously summarize $\bm{D}$. In this case, the assumption that the exposure mapping is structural typically rules out endogenous peer effects mediated by $\bm{A}$. 

We consider the following class of {\em $t'$-degenerate} exposure mappings.

\begin{myassump}{DEG}\label{adeg}
  For any $i\in\N_n$, there exists $\bm{\delta}_i \in \{0,1\}^{\abs{\N(i,K)}}$ such that, for any $\bm{d}\in\{0,1\}^n$, $f(\bm{d}_{\N(i,K)})=t'$ implies $Y_i(\bm{d}) = Y_i(\bm{\delta}_i, \bm{d}_{-\N(i,K)})$ a.s.
\end{myassump}

\noindent In other words, when $T_i=t'$, this pins down the treatment subvector on $i$'s $K$-neighborhood for the potential outcome. The assumption is a restriction on $f$ and also potentially on potential outcomes. The exposure mapping in the next example does not require any restrictions on $Y_i(\cdot)$ while the subsequent example does.

\begin{example}[Treated Neighbor Count]\label{etnc}
  Consider the treated neighbor count from \autoref{enc} with $t = (d,\eta)$ and $t' = (d',\eta')$ for some $d,d' \in \{0,1\}$ and $\eta,\eta' \in \mathbb{N} \cup \{0\}$. Then the exposure is $t'$-degenerate if $\eta'=0$ since $T_i=t'$ means all neighbors are untreated, so $Y_i(\bm{d}) = Y_i( (d', 0, \ldots, 0), \bm{d}_{-\N(i,1)} )$. This example requires no restrictions on potential outcomes. On the other hand, if $\eta' \in (0, \sum_{j=1}^n A_{ij})$, then $T_i=t'$ does not pin down which of $i$'s neighbors are treated, so $t'$-degeneracy does not hold. Thus the spirit of $t'$-degeneracy here is that the exposure value $t'$ constitutes a ``base case'' where all units in the $K$-neighborhood receive identical treatments.
\end{example}

While treated neighbor counts are covered by \autoref{tcount}, the result that follows imposes a different restriction on the assignment mechanism and allows for non-count exposures, a leading case of which is the following.

\begin{example}[Local Configuration]\label{eTiso}
  Let $\bm{A}_{\N(i,K)}$ be the subnetwork on $\N(i,K)$, that is $(A_{jk}\colon j,k \in \N(i,K))$. Restrict the population used in the exposure contrast to the subset of units $i$ such that $\bm{A}_{\N(i,K)} \cong \bm{a}$ for some network $\bm{a}$ where $\cong$ denotes graph isomorphism,\footnote{Define a {\em permutation} $\pi$ as a bijection on $\N_n$. Abusing notation, write $\pi(\bm{D}) = (D_{\pi(i)})_{i=1}^n$ and similarly $\pi(\bm{A}) = (A_{\pi(i)\pi(j)})_{i,j}$, which permutes the rows and columns of the matrix $\bm{A}$. If there exists a permutation $\pi$ such that $(\bm{D}_{\N(i,K)}, \bm{A}_{\N(i,K)}) = (\pi(\bm{\delta}),\pi(\bm{a}))$, then we write $(\bm{D}_{\N(i,K)}, \bm{A}_{\N(i,K)}) \cong (\bm{\delta},\bm{a})$.} and define
  \begin{equation*}
    f(\bm{d}_{\N(i,K)}) = \left\{ \begin{array}{ll} 1 & \text{if } (\bm{d}_{\N(i,K)}, \bm{A}_{\N(i,K)}) \cong (\bm{\delta},\bm{a}) \\ 0 & \text{if } (\bm{d}_{\N(i,K)}, \bm{A}_{\N(i,K)}) \cong (\bm{\delta}',\bm{a}). \end{array} \right.
  \end{equation*}

  \noindent Then $\tau(1,0)$ compares the subset of units with $K$-neighborhood subnetwork $\bm{a}$ and $K$-neighborhood treatment configuration $\bm{\delta}'$ vs.\ $\bm{\delta}$. This is the estimand studied in \S3.2 of \cite{auerbach2023local}, which they label a ``policy effect.'' 

  Now suppose potential outcomes are invariant with respect to treatment configuration permutations on the $K$-neighborhood in that
  \begin{multline}
    Y_i(\bm{d}_{\N(i,K)}, \bm{d}_{-\N(i,K)}) = Y_i(\bm{\delta}', \bm{d}_{-\N(i,K)}) \\ \text{a.s. for all}\quad \bm{d} \in \{0,1\}^n \quad\text{s.t.}\quad (\bm{d}_{\N(i,K)}, \bm{A}_{\N(i,K)}) \cong (\bm{\delta}',\bm{a}). \label{ypinvar}
  \end{multline}

  \noindent Then $f$ is $t'$-degenerate for $t'=0$, which is the $(\bm{\delta}',\bm{a})$ case. This condition imposes a nontrivial restriction on heterogeneity. For example if $i$ is part of an isolated intransitive triad $j \leftrightarrow i \leftrightarrow k$, then \eqref{ypinvar} says that $i$'s response is the same whether only $j$ or only $k$ is treated, regardless of their characteristics. This can be weakened to allow for heterogeneity in alters' observed characteristics by including covariates in the exposure mapping.
\end{example}

\begin{remark}\label{riso}
  \cite{auerbach2023local} maintain the permutation-invariance restriction \eqref{ypinvar} (first paragraph of their p.\ 180) in a setup where $K$ is large enough to encompass the entire network, meaning $\bm{A} = \bm{A}_{\N(i,K)}$ for all $i$. The reason they can choose large $K$ is that they assume observation of many independent small networks. In this case, estimation is straightforward in principle by aggregating across networks, and the exposure mapping is trivially structural. 

  Our work is relevant for small $K$, or low-dimensional exposure mappings, which are also the focus of most of the interference literature. As pointed out by \cite{auerbach2023local}, these run the risk of being non-structural, but the reason for this focus is that it is necessary to handle more challenging settings in which the data consists of a single large network. However, when $K$ does not encompass the full network, \eqref{ypinvar} is more restrictive since the exposure is not structural. \autoref{tani} below considers a weaker version of \eqref{ypinvar} where it only holds ``approximately'' on the $K$-neighborhood.
\end{remark}

%---------------------------------------------
\subsection{Unrestricted Interference}\label{sdeg}
%---------------------------------------------

Our first result imposes no restrictions on interference but requires the assignment mechanism to satisfy the following.

\begin{myassump}{$K$-CI}\label{aCI}
  For any $i\in\N_n$, $\bm{D}_{\N(i,K)} \indep \bm{D}_{-\N(i,K)} \mid \C_i$.
\end{myassump}

\noindent This states that each unit's $K$-neighborhood treatment assignment vector is independent of the remaining assignments conditional on the controls. It holds if $\{D_i\}_{i=1}^n$ is independently distributed conditional on $\C_i$ for any $i$, which is the case considered in \autoref{sCI}. It also holds in CRTs for which treatments are independent across clusters. 

Clustering corresponds to the special case in which $\bm{A}$ is block diagonal, so $\N(i,1)$ is the set of units in $i$'s cluster. Unlike previous theorems, we allow for arbitrary correlation between assignments within cluster. For instance, each cluster can be a network, and within each network, one can implement the correlated designs discussed in \autoref{sCI}. Unlike \autoref{tCRT}, the result encompasses a different class of exposure mappings, such as treated neighbor counts which have also been used in the CRT literature \citep{miguel2004worms,vazquez2023identification}.

\begin{theorem}\label{tdeg}
  Under Assumptions \ref{aunc} and \ref{adeg}, $\tau(t,t') = \tau^*(t,t') + \mathcal{B}$ where
  \begin{align*}
    &\tau^*(t,t') = \frac{1}{n} \sum_{i=1}^n \E\big[Y_i(\bm{D}) - Y_i(\bm{\delta}_i, \bm{D}_{-\N(i,K)}) \mid T_i=t, \C_i \big] \quad\text{and} \\
    &\mathcal{B} = \frac{1}{n} \sum_{i=1}^n \left( \E\big[Y_i(\bm{\delta}_i,\bm{D}_{-\N(i,K)}) \mid T_i=t, \C_i\big] - \E\big[Y_i(\bm{\delta}_i,\bm{D}_{-\N(i,K)}) \mid T_i=t', \C_i\big] \right).
  \end{align*}

  \noindent Under \autoref{aCI}, $\mathcal{B}=0$.
\end{theorem}

\noindent The causal estimand $\tau^*(t,t')$ is a convex average of unit-level effects of the form $Y_i(\bm{d}_{\N(i,K)}, \bm{d}_{-\N(i,K)}) - Y_i(\bm{\delta}_i, \bm{d}_{-\N(i,K)})$ which fix treatments outside the $K$-neighborhood while varying $K$-neighborhood treatments subject to the exposure mapping constraint $f(\bm{d}_{\N(i,K)})=t$. 

The decomposition $\tau^*(t,t') + \mathcal{B}$ has an omitted variable bias interpretation. The first term $\tau^*(t,t')$ is the effect of variation in the ``regressor'' $\bm{D}_{\N(i,K)}$ induced by the exposure mapping. The bias $\mathcal{B}$ is nonzero if the $K$-neighborhood exposure is correlated with the ``omitted variable'' $\bm{D}_{-\N(i,K)}$. Unconfoundedness alone provides no control over $\mathcal{B}$. Theorem 4 of \cite{sobel2006randomized} and Theorem 1 of \cite{vazquez2023identification} provide analogous decompositions, but their results are limited to the case of $T_i=D_i$.

%---------------------------------------------
\subsection{Unrestricted Assignment Mechanism}\label{stselect}
%---------------------------------------------

The remaining results impose no restriction on the assignment mechanism other than unconfoundedness. The first result requires higher-order spillovers, meaning those induced by units beyond the ego's $K$-neighborhood, to be uniformly smaller than $K$-neighborhood spillovers. We refer to this as ``neighborhood-centric interference.''

\begin{myassump}{$K$-NCI}\label{anp}
  $\Delta_K > \Psi_K$ where
  \begin{align*}
    \Delta_K &= \min\big\{ \abs{Y_i(\bm{d}_{\N(i,K)}, \bm{d}_{-\N(i,K)}'') - Y_i(\bm{d}_{\N(i,K)}', \bm{d}_{-\N(i,K)}'')} \big\}, \\
    \Psi_K &= \max\big\{ \abs{Y_i(\bm{d}_{\N(i,K)}'', \bm{d}_{-\N(i,K)}) - Y_i(\bm{d}_{\N(i,K)}'', \bm{d}_{-\N(i,K)}')} \big\}, 
  \end{align*}

  \noindent and the max and min are taken over $i \in \N_n$ and $\bm{d},\bm{d}',\bm{d}'' \in \{0,1\}^n$.
\end{myassump}

\noindent The term $\Delta_K$ is the smallest $K$-neighborhood spillover effect across all units, while $\Psi_K$ is the largest higher-order spillover effect from beyond the $K$-neighborhood.  

\begin{example}
  In the case of $K=0$, $\Delta_K$ is the smallest direct effect of the treatment over all units, so $K$-NCI holds if direct effects uniformly dominate spillover effects in magnitude. This is relevant for settings in which direct effects are typically larger than spillover effects, for instance online experiments \citep{viviano2023causal,yuan2021causal}. In the context of vaccines, the spillover effect from reduced community transmission is often smaller than the effect of being directly vaccinated.
\end{example}

\begin{example}
  Several papers assume that potential outcomes only depend on treatments within a $K$-neighborhood but without imposing a particular exposure mapping:
  \begin{equation}
    Y_i(\bm{d}) = Y_i(\bm{d}') \quad\text{for all}\quad \bm{d},\bm{d}'\in\{0,1\}^n \quad\text{such that}\quad \bm{d}_{\N(i,K)} = \bm{d}_{\N(i,K)}' \label{ani}
  \end{equation}

  \noindent \citep[e.g.][]{ugander2013graph,viviano2023causal}. This implies $K$-NCI since $\Psi_K=0$. If $\bm{A}$ is block-diagonal, with each block representing a fully connected cluster, then partial interference corresponds to \eqref{ani} for any $K\geq 1$. $K$-NCI allows for violations of partial interference, so long as cross-cluster interference $\Psi_K$ is uniformly dominated by within-cluster interference $\Delta_K$. 
\end{example}

\begin{theorem}\label{tnci}
  Suppose $Y_i(\bm{d}) - Y_i(\bm{d}') \geq (\leq) \, 0$ for all $i \in \N_n$ and $\bm{d},\bm{d}' \in \{0,1\}^n$ such that $\bm{d}_{-\N(i,K)} = \bm{d}_{-\N(i,K)}'$ and $f(i,\bm{d})=t$ and $f(i,\bm{d}')=t'$. Then under Assumptions \ref{aunc}, \ref{adeg}, and \ref{anp}, $\tau(t,t') \geq (\leq) \, 0$.
\end{theorem}

\noindent The unit-level effects in the theorem are those in the convex average $\tau^*(t,t')$, so if these possess the same sign, then so does $\tau^*(t,t')$. Under \autoref{anp}, $\mathcal{B}$ is smaller in magnitude than $\tau^*(t,t')$, so $\tau(t,t') = \tau^*(t,t') + \mathcal{B}$ preserves the sign of $\tau^*(t,t')$ and therefore those of the unit-level effects. \cite{sobel2006randomized} previously noted in the context of a particular class of experiments that when $T_i=D_i$, $\tau(t,t')$ is not subject to sign reversals when direct effects dominate spillover effects. \autoref{tnci} generalizes this to $t'$-degenerate exposure mappings and arbitrary designs.

The result is motivated by the omitted variable bias interpretation of \autoref{tdeg}. By definition, $K$-neighborhood exposures directly manipulate $\bm{D}_{\N(i,K)}$, but since treatments are correlated, they also induce variation in $\bm{D}_{-\N(i,K)}$, which generates bias $\mathcal{B}$. The bias involves spillovers beyond the $K$-neighborhood. Under $K$-NCI, it is smaller in magnitude than the causal estimand $\tau^*(t,t')$, so the sign of the latter dominates.\footnote{I thank a referee for comments that inspired this result.} 

Our last result considers $K$-neighborhood exposure mappings with $K$ chosen relatively large, as may be the case in \autoref{eTiso}. We require interference between units to decay with their distance. The main idea is that larger $K$ implies that the exposure mapping captures variation in the treatment subvector within a larger radius. Since interference beyond this radius is relatively small, the exposure is ``approximately'' structural, so $\tau(t,t')$ should approximate a quantity that has a causal interpretation. This formalizes some of the discussion in \S4 of \cite{auerbach2024discussion}.

We consider the \cite{leung2024graph} model \eqref{YD}. For any $S \subseteq \mathcal{N}_n$, let $\bm{X}_S = (X_i)_{i\in S}$, and similarly define $\bm{\varepsilon}_S$ and $\bm{\nu}_S$. The following ``approximate neighborhood interference'' condition due to \cite{leung2024graph} formalizes the idea of interference decaying to zero as path distance diverges.

\begin{myassump}{ANI}\label{aani}
  There exists $\gamma \colon \R_+\rightarrow \R_+$ such that $\gamma(s) \stackrel{s\rightarrow\infty}\longrightarrow 0$ and
  \begin{multline*}
    \max_{i\in\N_n} \E\big[\lvert g_n(i, \bm{D}, \bm{X}, \bm{A}, \bm{\varepsilon}) \\ - g_{\abs{\N(i,s)}}(i, \bm{D}_{\N(i,s)}, \bm{X}_{\N(i,s)}, \bm{A}_{\N(i,s)}, \bm{\varepsilon}_{\N(i,s)}) \rvert \mid \bm{D}, \bm{X}, \bm{A} \big] \leq \gamma(s).
  \end{multline*}
\end{myassump}

\noindent To understand the inequality, first recall from \eqref{YD} that $g_n(i, \bm{D}, \bm{X}, \bm{A}, \bm{\varepsilon})$ is $i$'s observed outcome $Y_i$. We interpret $g_{\abs{\N(i,s)}}(i,\dots)$ as $i$'s outcome under a counterfactual ``$s$-neighborhood model'' in which the primitives and treatments are fixed at their realizations, units external to the $s$-neighborhood are excluded from the model, and the remaining units interact according to the reduced-form model $g_{\abs{\N(i,s)}}(\cdot)$. ANI bounds the difference between $i$'s realized outcome and counterfactual $s$-neighborhood outcome by $\gamma(s)$, which is required to decay with the radius $s$. If the rate of decay is faster, then $Y_i$ is well-approximated by a model with only units in $\N(i,s)$ for smaller $s$, formalizing the idea that units distant from $i$ interfere less with $i$.

Let $\tilde{Y}_i^K(\bm{d}_i) = g_{\abs{\N(i,K)}}(i, \bm{d}_i, \bm{X}_{\N(i,K)}, \bm{A}_{\N(i,K)}, \bm{\varepsilon}_{\N(i,K)})$, the potential outcome under the counterfactual $K$-neighborhood model,
and
\begin{equation*}
  \tau^*_K(t,t') = \frac{1}{n} \sum_{i=1}^n \big( \E[\tilde{Y}_i^K(\bm{D}_{\N(i,K)}) - \tilde{Y}_i^K(\bm{\delta}_i) \mid T_i=t, \C_i] \big).
\end{equation*}

\noindent This is similar to $\tau^*(t,t')$ in that it is a convex average of unit-level effects, except it uses potential outcomes from the counterfactual $K$-neighborhood model. 

\begin{theorem}\label{tani}
  Consider model \eqref{YD} with controls \eqref{confounders}. Suppose the exposure mapping is $t'$-degenerate under the counterfactual $K$-neighborhood model in that, for all $i$, there exist some $\bm{\delta}_i \in \{0,1\}^{\abs{\N(i,K)}}$ such that, for any $\bm{d}_{\N(i,K)} \in \{0,1\}^{\abs{\N(i,K)}}$, $f(\bm{d}_{\N(i,K)})=t'$ implies 
  \begin{equation}
    \tilde{Y}_i^K(\bm{d}_{\N(i,K)}) = \tilde{Y}_i^K(\bm{\delta}_i) \quad \text{a.s.} \label{adegweak}
  \end{equation}

  \noindent Under Assumptions \ref{aunc} and \ref{aani}, $\abs{\tau(t,t') - \tau^*_K(t,t')} \leq 2\gamma(K) \rightarrow 0$ as $K\rightarrow\infty$.
\end{theorem}

\noindent Equation \eqref{adegweak} weakens \autoref{adeg} to hold only for the counterfactual model in which potential outcomes only depend on the $K$-neighborhood. In \autoref{eTiso}, this holds if we replace \eqref{ypinvar} with the weaker requirement
\begin{equation*}
  \tilde{Y}_i(\bm{d}_{\N(i,K)}) = \tilde{Y}_i(\bm{\delta}') \quad\text{a.s.\ for all}\quad \bm{d}_{\N(i,K)} \quad\text{s.t.}\quad (\bm{d}_{\N(i,K)}, \bm{A}_{\N(i,K)}) \cong (\bm{\delta}',\bm{a}). 
\end{equation*}

\noindent Per the discussion in \autoref{riso}, this is equivalent to the restriction employed by \cite{auerbach2023local} since the $K$-neighborhood is the ``entire network'' under the counterfactual model.

Under \eqref{ani}, ANI holds with $\gamma(s)=0$ for all $s\geq K$, in which case $\tau(t,t') = \tau_K^*(t,t')$ which has an exact causal interpretation. \cite{leung2022causal} shows that ANI can be satisfied by well-known models of social interactions with exponentially decaying $\gamma(s)$. In this case, choosing $K$ to be logarithmic in $n$ can ensure that the bias $\abs{\tau(t,t') - \tau^*_K(t,t')}$ is order $n^{-c}$ for some $c>0$.

%----------------------------------------------------------------------
\section{Conclusion}\label{sconclude}
%----------------------------------------------------------------------

In settings with interference, researchers often report exposure contrasts to summarize treatment and spillover effects. A common example is to regress an outcome on own treatment assignment and the number or share of treated neighbors. Researchers typically interpret the respective coefficients as ``direct'' and ``spillover'' effects, but we show that this interpretation is not generally valid. The exposure contrast can have the opposite sign of the unit-level effects of interest even if treatment assignment is unconfounded.

Eliminating sign reversals requires restricting either interference or correlation in treatment assignments across units. The literature typically restricts interference by assuming that it is entirely mediated by a known, low-dimensional exposure mapping. In our view, typical choices, such as the number of treated neighbors, function more as statistics of convenience and are unlikely to fully determine interference, particularly in the presence of endogenous peer effects. We propose alternative assumptions that are substantially weaker and rule out sign reversals. 

Our first result considers assignments satisfying a conditional independence condition. We show that this can be satisfied by stratified experiments and selection models with peer effects. Our second result concerns cluster-randomized trials, and we show that standard estimands can be written as convex averages of unit-level effects without imposing any restrictions on spillovers within or across clusters. Finally, we consider arbitrary unconfounded assignment mechanisms and show that sign reversals can be avoided under various restrictions on interference that allow for endogenous peer effects.

\appendix
\numberwithin{equation}{section} % include section number in equation numbering

%----------------------------------------------------------------------
\section{Sign Preservation Criteria}\label{ssignp}
%----------------------------------------------------------------------

This section discusses the definition of sign preservation criteria for exposure contrasts. These rule out undesirable sign reversals, providing a minimal formal sense in which $\tau(t,t')$ is ``causal.'' Under SUTVA, there is only one natural definition of sign preservation: $n^{-1} \sum_{i=1}^n (\E[Y_i \mid D_i=1] - \E[Y_i \mid D_i=0]) \geq 0$ ($\leq 0$) if $Y_i(1) - Y_i(0) \geq 0$ ($\leq 0$) for all $i$ \citep{blandhol2022tsls,bugni2023decomposition}. We will see that, under interference, there are many possible definitions, and which is relevant depends on the exposure mapping and the unit-level comparisons of interest.

In response to results from an earlier version of this paper, \cite{savje2024rejoinder} proposes a sign preservation criterion for arbitrary exposure mappings and shows that, under a randomized control trial, $\tau(t,t')$ satisfies it without any further restrictions. This was intended to refute our claim that additional restrictions are required to ensure that $\tau(t,t')$ avoids sign reversals. We discuss the problem with this result and how it relates to our theorems.

%---------------------------------------------
\subsection{Treatment Sign Preservation}
%---------------------------------------------

Let us first consider the familiar case where the exposure mapping and contrast are $T_i=D_i$ and $\tau(1,0)$. We will extend the SUTVA sign preservation criterion to the case with interference. The unit-level treatment effects of interest are now
\begin{equation*}
  \mathcal{T}_i^\text{TSP}(1,0) = \big\{ Y_i(1,\bm{d}_{-i}) - Y_i(0,\bm{d}_{-i}) \colon \bm{d}_{-i} \in \{0,1\}^{n-1} \big\}.
\end{equation*}

\noindent The following criterion provides a minimal sense in which $\tau(1,0)$ is informative for these effects.

\begin{secdefinition}[TSP]\label{TSP}
  $\tau(1,0)$ is {\em treatment sign preserving (TSP)} if $\min_i \min \mathcal{T}_i^\text{TSP}(1,0) \geq 0$ implies $\tau(1,0) \geq 0$ and $\max_i \max \mathcal{T}_i^\text{TSP}(1,0) \leq 0$ implies $\tau(1,0) \leq 0$.
\end{secdefinition}

\noindent That is, $\tau(1,0)$ is positive (negative) if all unit-level treatment effects are positive (negative). The next result shows that unconfoundedness alone is insufficient to ensure the TSP property. This is not a new insight in the literature; \cite{eck2022randomization} and \cite{sobel2006randomized} show that $\tau(1,0)$ is not informative for the direct effect of treatment under interference. 

\begin{secproposition}\label{ptsp}
  There exist potential outcomes and an assignment mechanism satisfying \autoref{aunc} such that $\tau(1,0)$ is not treatment sign preserving.
\end{secproposition}
\begin{proof}
  Suppose $T_i=D_i$. Let $n=2$, so that we may write $Y_i(\bm{d}) = Y_i(d_1,d_2)$ where $d_1$ is unit 1's counterfactual assignment and $d_2$ is unit 2's. Consider the potential outcomes

  \hfill
  \begin{center}
    \begin{tabular}{lll}
      $Y_1(0,0) = Y_2(0,0) = 0$ & & $\prob(\bm{D}=(0,0) \mid D_1=0) = \prob(\bm{D}=(0,0) \mid D_2=0) = p_1$ \\
      $Y_1(1,0) = Y_2(0,1) = 1$ & & $\prob(\bm{D}=(1,0) \mid D_1=1) = \prob(\bm{D}=(0,1) \mid D_2=1) = p_2$ \\
      $Y_1(0,1) = Y_2(1,0) = 2$ & & $\prob(\bm{D}=(0,1) \mid D_1=0) = \prob(\bm{D}=(1,0) \mid D_2=0) = p_3$ \\
      $Y_1(1,1) = Y_2(1,1) = 3$ & & $\prob(\bm{D}=(1,1) \mid D_1=1) = \prob(\bm{D}=(1,1) \mid D_2=1) = p_4$
    \end{tabular}
  \end{center}

  \hfill

  \noindent The restriction to $n=2$ is for simplicity, and the example is easily scaled up by considering a large population of identical dyads. Notice there is no heterogeneity across the two units, so
  \begin{equation*}
    \tau(1,0) = \E[Y_1 \mid D_1=1] - \E[Y_1 \mid D_1 = 0] = 3p_4 + p_2 - 2p_3.
  \end{equation*}

  \noindent Unit-level treatment effects are positive since $Y_1(1,d_2) - Y_1(0,d_2) = Y_2(d_1,1) - Y_2(d_1,0) = 1$ for any $d_1,d_2 \in \{0,1\}$, but it is straightforward to construct assignment mechanisms such that $\tau(1,0) < 0$. For example, consider complete randomization where half of the units (one in each dyad) are allocated to treatment: $\prob(\bm{D}=(1,0)) = \prob(\bm{D}=(0,1)) = 0.5$. Then $(p_1, p_2, p_3, p_4) = (0, 1, 1, 0)$, so $\tau(1,0) = -1$, which violates TSP. 
\end{proof}

%---------------------------------------------
\subsection{General Sign Preservation}
%---------------------------------------------

\cite{savje2024rejoinder} proposes the following sign preservation criterion for arbitrary exposure mappings.

\begin{secdefinition}[GSP]\label{dGSP}
  Define the comparison set
  \begin{equation*}
    \mathcal{T}_i^\text{GSP}(t,t') = \big\{ Y_i(\bm{d}) - Y_i(\bm{d}') \colon \bm{d},\bm{d}' \in \{0,1\}^n, f(i,\bm{d})=t, f(i,\bm{d}')=t' \big\}.
  \end{equation*}

  \noindent We say $\tau(t,t')$ is {\em general sign preserving (GSP)} if $\min_i \min \mathcal{T}_i^\text{GSP}(t,t') \geq 0$ implies $\tau(t,t') \geq 0$ and $\max_i \max \mathcal{T}_i^\text{GSP}(t,t') \leq 0$ implies $\tau(t,t') \leq 0$.
\end{secdefinition}

\noindent The set $\mathcal{T}_i^\text{GSP}(t,t')$ contains the unit-level effects of varying the entire treatment assignment vector subject to the constraints that $f(i,\bm{d})=t$ and $f(i,\bm{d}')=t'$. 

The next result shows that $\tau(t,t')$ is always GSP under unconfoundedness. This does not contradict \autoref{ptsp}, as we discuss below.

\begin{secproposition}[\cite{savje2024rejoinder}, Proposition 1]\label{pGSP}
  Under \autoref{aunc}, $\tau(t,t')$ is general sign preserving for {\em any} exposure mapping.
\end{secproposition}
\begin{proof}
  We provide a simpler proof using the basic fact that a difference of convex averages can always be written as a convex average of differences. Abbreviating $\sigma_{i,t}(\bm{d}) \equiv \prob(\bm{D}=\bm{d} \mid T_i=t, \C_i)$,
  \begin{align*}
    \frac1n &\sum_{i=1}^n \big( \E[Y_i \mid T_i=t, \C_i] - \E[Y_i \mid T_i=t', \C_i] \big) \\
	    &= \frac1n \sum_{i=1}^n \sum_{\bm{d}\in\{0,1\}^n} \sum_{\bm{d}'\in\{0,1\}^n} \big( \E[Y_i(\bm{d}) \mid \bm{D}=\bm{d}, \C_i] - \E[Y_i(\bm{d}') \mid \bm{D}=\bm{d}', \C_i] \big) \sigma_{i,t}(\bm{d}) \sigma_{i,t'}(\bm{d}') \\
	    &= \frac1n \sum_{i=1}^n \sum_{\bm{d}\in\{0,1\}^n} \sum_{\bm{d}'\in\{0,1\}^n} \E[Y_i(\bm{d}) - Y_i(\bm{d}') \mid \C_i] \sigma_{i,t}(\bm{d}) \sigma_{i,t'}(\bm{d}')
  \end{align*}

  \noindent the last line using \autoref{aunc}. This is a convex average of elements in $\mathcal{T}_i^\text{GSP}(t,t')$ over all $i$, so GSP follows.
\end{proof}

The problem with this result is that the set $\mathcal{T}_i^\text{GSP}(t,t')$ contains undesirable comparisons. This makes the set too large, the requirement $\min_i \min \mathcal{T}_i^\text{GSP}(t,t') \geq 0$ ($\leq 0$) too demanding, and GSP potentially vacuous. To see this, specialize to the TSP case of $T_i=D_i$, $t=1$, and $t'=0$. Then 
\begin{equation}
  \mathcal{T}_i^\text{GSP}(t,t') = \left\{ Y_i(1,\bm{d}_{-i}) - Y_i(0,\bm{d}_{-i}') \colon \bm{d},\bm{d}' \in \{0,1\}^n \right\}. \label{wrong}
\end{equation}

\noindent Unlike $\mathcal{T}_i^\text{TSP}(1,0)$, this is missing the constraint $\bm{d}_{-i}=\bm{d}_{-i}'$, so $Y_i(1,\bm{d}_{-i}) - Y_i(0,\bm{d}_{-i}')$ is not a treatment effect. In the example in the proof of \autoref{ptsp}, GSP is vacuous because the elements in \eqref{wrong} do not all share the same sign ($\min_i \min \mathcal{T}_i^\text{GSP}(t,t') \geq 0$ is too demanding), so the criterion ends up imposing no restrictions on $\tau(t,t')$. In contrast, $\mathcal{T}_i^\text{TSP}(1,0)$ only contains treatment effects, which are the comparisons relevant to the exposure $T_i=D_i$. This makes TSP a stronger, more discriminating criterion, hence the negative result in \autoref{ptsp}. 

The relevant sign preservation criterion thus depends on the exposure mapping and unit-level effects of interest. To repair the GSP criterion, $\mathcal{T}_i^\text{GSP}(t,t')$ should be replaced with the appropriate unit-level comparisons. For count exposures, \autoref{tcount} suggests
\begin{equation*}
  \mathcal{T}_i(t,t') = \big\{ Y_i(\bm{d}) - Y_i(\bm{d}') \colon \bm{d},\bm{d}'\in \{0,1\}^n, \bm{d} \geq \bm{d}', f(i,\bm{d})=t, f(i,\bm{d}')=t' \big\},
\end{equation*}

\noindent which adds the constraint $\bm{d} \geq \bm{d}'$. For $K$-neighborhood exposures, \autoref{tdeg}--\ref{tani} suggest
\begin{multline*}
  \mathcal{T}_i(t,t') = \big\{ Y_i(\bm{d}) - Y_i(\bm{d}') \colon \bm{d},\bm{d}' \in \{0,1\}^n, \bm{d}_{-\N(i,K)} = \bm{d}_{-\N(i,K)}', \\ f(i,\bm{d})=t, f(i,\bm{d}')=t' \big\}.
\end{multline*} 

\noindent which adds the constraint that $\bm{d}_{-\N(i,K)} = \bm{d}_{-\N(i,K)}'$. The special case of $K=0$ corresponds to $\mathcal{T}_i^\text{TSP}(1,0)$. 

%----------------------------------------------------------------------
\section{Supporting Lemmas}
%----------------------------------------------------------------------

\begin{seclemma}\label{lsd}
  For random vectors $\bm{X}^{(1)}, \bm{X}^{(2)}$ of the same dimension, $\bm{X}^{(1)}$ first-order stochastically dominates $\bm{X}^{(2)}$ if and only if $\E[\phi(\bm{X}^{(1)})] \geq \E[\phi(\bm{X}^{(2)})]$ for all increasing functions $\phi$ for which the expectations exist.
\end{seclemma}
\begin{proof}
  See (6.B.4) of \cite{shaked2007stochastic}.
\end{proof}

\begin{seclemma}\label{lstrass}
  Let $\bm{X}_1, \ldots, \bm{X}_n$ be random vectors of the same dimension such that $\bm{X}_i$ first-order stochastically dominates $\bm{X}_j$ for all $i > j$. Then there exist $\bm{X}_1^*, \ldots, \bm{X}_n^*$ defined on the same probability space such that $\bm{X}_i^* \stackrel{d}= \bm{X}_i$ for all $i$ and $\bm{X}_i^* \geq \bm{X}_j^*$ a.s.\ for all $i > j$.
\end{seclemma}
\begin{proof}
  See the paragraph after Theorem 6.B.2 of \cite{shaked2007stochastic}. 
\end{proof}

\begin{secdefinition}\label{dlc}
  The PMF $p(\cdot)$ of a discrete random variable is {\em log-concave} if $p(t)^2 \geq p(t-1)p(t+1)$ for all $t$.
\end{secdefinition}

\begin{seclemma}\label{lefron}
  Suppose $X_1,\dots,X_n$ are independent discrete random variables with log-concave PMFs. Then for any componentwise nondecreasing $\Phi\colon \R^n \rightarrow \R$,
  \begin{equation*}
    \E\left[ \Phi(X_1, \ldots, X_n) \,\bigg|\, \sum_{i=1}^n X_i = t \right]
  \end{equation*}

  \noindent is nondecreasing in $t$.
\end{seclemma}
\begin{proof}
  See Remark 6.2 of \cite{saumard2014log}.
\end{proof}

\begin{seclemma}\label{lmc}
  Let $\bm{X} = (X_1,\dots,X_n)$ be a vector of independent random variables with $X_i \sim \mathrm{Bernoulli}(p_i)$ for $p_i \in (0,1)$ and $S = \sum_{i=1}^n X_i$. Let $I_- \leq I_+$ be possibly unbounded intervals (recall the notation prior to \autoref{aCT}). There exists a monotone coupling $\bm{X}^{(-)} \leq \bm{X}^{(+)}$ a.s.\ such that $\bm{X}^{(a)} \stackrel{d}= (\bm{X} \mid S \in I_a)$ for $a \in \{-,+\}$.
\end{seclemma}
\begin{proof}
  We begin by constructing a monotone coupling $\bm{X}^{(0)} \leq \ldots \leq \bm{X}^{(n)}$ a.s.\ such that $\bm{X}^{(s)} \stackrel{d}= (\bm{X} \mid S=s)$ for all $s$. By Example 1.3(3) of \cite{johnson2006preservation}, the PMF of a Bernoulli random variable is log-concave, so by \autoref{lefron}, $\E\left[ \Phi(\bm{X}) \mid S = s \right]$ is nondecreasing in $s$ for any componentwise nondecreasing $\Phi$. Then by \autoref{lsd}, for any $t\geq t'$, the distribution of $\bm{X} \mid S = t$ stochastically dominates that of $\bm{X} \mid S = t'$. Existence of the monotone coupling follows from \autoref{lstrass}.

  For any possibly unbounded interval $I$, let $R_I \stackrel{d}= (S \mid S \in I)$ be independent of the monotone coupling $(\bm{X}^{(0)}, \ldots, \bm{X}^{(n)})$ constructed in step 1. Then $\bm{X}^{(R_I)} \stackrel{d}= (\bm{X} \mid S \in I)$ since
  \begin{equation*}
    \prob(\bm{X} = \bm{x} \mid S \in I) = \sum_{s\in I} \prob(\bm{X} = \bm{x} \mid S = s) \prob(S = s \mid S \in I).
  \end{equation*}

  Now draw $R_- \stackrel{d}= (S \mid S \in I_-)$ and $R_+ \stackrel{d}= (S \mid S \in I_+)$ independently of the monotone coupling. Since $I_- \leq I_+$, $R_-\leq R_+$ a.s. Hence, $\bm{X}^{(-)} \equiv \bm{X}^{(R_-)} \leq \bm{X}^{(+)} \equiv \bm{X}^{(R_+)}$ a.s., and they have the claimed marginal distributions by the previous paragraph.
\end{proof}

%----------------------------------------------------------------------
\section{Proofs}\label{sproofs}
%----------------------------------------------------------------------

\begin{proof}[Proof of \autoref{tcount}]
  Let $\bm{D}_{i,-}$ be the subvector of $\bm{D}$ restricted to $\N_n\backslash (B_{i1} \cup \ldots \cup B_{id_t})$. Let $\bm{D}_{i,t}$ be a draw from $p_{i,t}(\cdot \mid c)$ and similarly for $\bm{D}_{i,t'}$. Using \autoref{aCT}, partition
  \begin{equation*}
    \bm{D}_{i,t} = (\bm{D}_{B_{i1}}^t, \ldots, \bm{D}_{B_{id_t}}^t, \bm{D}_{i,-}^t) \quad\text{and}\quad \bm{D}_{i,t'} = (\bm{D}_{B_{i1}}^{t'}, \ldots, \bm{D}_{B_{id_t}}^{t'}, \bm{D}_{i,-}^{t'})
  \end{equation*}

  \noindent where $\bm{D}_{B_{ir}}^t$ is the subvector of $\bm{D}_{i,t}$ restricted to units in $B_{ir}$, while $\bm{D}_{i,-}^t$ is the subvector restricted to units in $\N_n\backslash (B_{i1} \cup \ldots \cup B_{id_t})$. By \autoref{aCI0}, $\bm{D}_{B_{i1}}^t \indep \ldots \indep \bm{D}_{B_{id_t}}^t \indep \bm{D}_{i,-}^t$ conditional on $\C_i=c$ and similarly for $t'$. 

  Using \autoref{lmc} and the fact that $I_{ir}^{(t)} \geq I_{ir}^{(t')}$ from \autoref{aCT}, we can construct monotone couplings $\bm{D}_{B_{ir}}^{t,*} \geq \bm{D}_{B_{ir}}^{t',*}$ a.s.\ for each pair $(\bm{D}_{B_{ir}}^t, \bm{D}_{B_{ir}}^{t'})$, $r=1,\ldots,d_t$. Furthermore, since $(\bm{D}_{B_{i1}}^t, \ldots, \bm{D}_{B_{id_t}}^t)$ and $(\bm{D}_{B_{i1}}^{t'}, \ldots, \bm{D}_{B_{id_t}}^{t'})$ are independent of $\bm{D}_{i,-}$ and $Y_i(\cdot)$ conditional on $\C_i=c$ by Assumptions \ref{aunc} and \ref{aCI0}, we can extend the probability space on which the latter two are defined to construct the couplings, $Y_i(\cdot)$, and $\bm{D}_{i,-}$ all on the same probability space.

  Define for $s \in \{t,t'\}$
  \begin{equation*}
    \bm{D}_{i,s}^* = (\bm{D}_{B_{i1}}^{s,*}, \ldots, \bm{D}_{B_{id_t}}^{s,*}, \bm{D}_{i,-}).
  \end{equation*}

  \noindent Since $\bm{D}_{i,-}$, $\bm{D}_{i,-}^t$, and $\bm{D}_{i,-}^{t'}$ have the same distributions conditional on $\C_i=c$, so do $\bm{D}_{i,s}^*$ and $\bm{D}_{i,s}$ by \autoref{aCI0}. We have therefore constructed $\bm{D}_{i,t}^* \geq \bm{D}_{i,t'}^*$ a.s.\ such that
  \begin{multline*}
    \E[Y_i(\bm{D}) \mid T_i=t, \C_i=c] - \E[Y_i(\bm{D}) \mid T_i=t', \C_i=c] \\ = \E\big[ Y_i(\bm{D}_{i,t}^*) - Y_i(\bm{D}_{i,t'}^*) \mid \C_i=c \big].
  \end{multline*}
\end{proof}

%-----------------
\begin{proof}[Proof of \autoref{tCRT}]
  Fix any $i\in\N_n$. By \autoref{aCRT}, it suffices to construct $\bm{D}_{(i),t}^* \stackrel{a.s.}\geq \bm{D}_{(i),t'}^*$ such that, for any $\bm{d}_{(-i)}$,
  \begin{multline}
    \E\big[ Y_i(\bm{D}_{(i)}, \bm{d}_{(-i)}) \mid T_i=t \big] - \E\big[ Y_i(\bm{D}_{(i)}, \bm{d}_{(-i)}) \mid T_i=t' \big] \\ = \E\big[ Y_i(\bm{D}_{(i),t}^*, \bm{d}_{(-i)}) - Y_i(\bm{D}_{(i),t'}^*, \bm{d}_{(-i)}) \big]. \label{gajowi}
  \end{multline}

  \bigskip
  \noindent {\bf Indirect and total effect.} Without loss of generality, let $i=1$ and $C_{(1)} = \{1, \ldots, \gamma\}$. Let $\{U_j\colon j=2, \ldots, \gamma\} \stackrel{iid}\sim \U([0,1])$ be independent of potential outcomes. Recall that the event $T_1=t'$ means conditioning on the saturation level $S_1$ being $p' \in \mathcal{P}$. Construct $\bm{D}_{(1),t'}^*$ by setting unit 1's treatment to $0$ in the case of the total effect and $d\in\{0,1\}$ in the case of the indirect effect and assigning units $j=2,\ldots,\gamma$ to treatment if and only if $U_j \leq p'$. By \autoref{aCRT}, $\bm{D}_{(1),t'}^*$ has the same distribution as $\bm{D}_{(1)} \mid T_1=t'$. 

  Recall that the event $T_1=t$ means conditioning on the saturation level $S_1$ being $p \in \mathcal{P}$ for $p\geq p'$. Construct $\bm{D}_{(1),t}^*$ by setting unit 1's treatment to $1$ in the case of the total effect and $d$ in the case of the indirect effect and assigning units $j=2,\ldots,\gamma$ to treatment if and only if $U_j \leq p$. By \autoref{aCRT}, $\bm{D}_{(1),t}^*$ has the same distribution as $\bm{D}_{(1)} \mid T_1=t$. By construction, $\bm{D}_{(1),t}^* \geq \bm{D}_{(1),t'}^*$ a.s., so \eqref{gajowi} holds. 

  \bigskip
  \noindent {\bf Overall effect.} Without loss of generality, let $i=1$ and $C_{(1)} = \{1, \ldots, \gamma\}$. Let $\{U_j\colon j=1, \ldots, \gamma\} \stackrel{iid}\sim \U([0,1])$. Construct $\bm{D}_{(1),t'}^*$ by assigning units $j=1,\ldots,\gamma$ to treatment if and only if $U_j \leq p'$. By \autoref{aCRT}, this has the same distribution as $\bm{D}_{(1)} \mid T_1=t'$. 

  Construct $\bm{D}_{(1),t}^*$ by assigning units $j=1,\ldots,\gamma$ to treatment if and only if $U_j \leq p$. By \autoref{aCRT}, $\bm{D}_{(1),t}^*$ has the same distribution as $\bm{D}_{(1)} \mid T_1=t$. By construction, $\bm{D}_{(1),t}^* \geq \bm{D}_{(1),t'}^*$ a.s., so \eqref{gajowi} holds. 
\end{proof}

%-----------------
\begin{proof}[Proof of \autoref{tdeg}]
  By \autoref{aunc},
  \begin{multline*}
    \tau(t,t') = \frac{1}{n} \sum_{i=1}^n \bigg( \sum_{\bm{d} \in \{0,1\}^n} \E[Y_i(\bm{d}) \mid \C_i] \prob(\bm{D}=\bm{d} \mid T_i=t, \C_i) \\ - \sum_{\bm{d}' \in \{0,1\}^n} \E[Y_i(\bm{d}') \mid \C_i] \prob(\bm{D}=\bm{d}' \mid T_i=t', \C_i) \bigg). 
  \end{multline*}

  \noindent By \autoref{adeg}, this equals
  \begin{multline*}
    \frac{1}{n} \sum_{i=1}^n \bigg( \sum_{\bm{d} \in \{0,1\}^n} \E[Y_i(\bm{d}) \mid \C_i] \,\prob(\bm{D}=\bm{d} \mid T_i=t, \C_i) \\ - \sum_{\bm{d}' \in \{0,1\}^n} \E[Y_i(\bm{\delta}_i,\bm{d}_{-\N(i,K)}') \mid \C_i] \,\prob(\bm{D}=\bm{d}' \mid T_i=t', \C_i) \bigg).
  \end{multline*}

  \noindent Add and subtract
  \begin{equation*}
    \frac{1}{n} \sum_{i=1}^n \sum_{\bm{d}' \in \{0,1\}^n} \E[Y_i(\bm{\delta}_i,\bm{d}_{-\N(i,K)}') \mid \C_i] \,\prob(\bm{D}=\bm{d}' \mid T_i=t, \C_i),
  \end{equation*}

  \noindent and the result equals $\tau^*(t,t') + \mathcal{B}$. By \autoref{aCI},
  \begin{multline*}
    \mathcal{B} = \frac{1}{n} \sum_{i=1}^n \bigg( \sum_{\bm{d}_{-\N(i,K)}} \E\big[Y_i(\bm{\delta}_i,\bm{d}_{-\N(i,K)}) \mid \C_i\big] \prob(\bm{D}_{-\N(i,K)}=\bm{d}_{-\N(i,K)} \mid \C_i)\\ \times \,\bigg( \sum_{\bm{d}_{\N(i,K)}} \prob(\bm{D}_{\N(i,K)}=\bm{d}_{\N(i,K)} \mid T_i=t, \C_i) \\ - \sum_{\bm{d}_{\N(i,K)}} \prob(\bm{D}_{\N(i,K)}=\bm{d}_{\N(i,K)} \mid T_i=t', \C_i) \bigg) \bigg).
  \end{multline*}

  \noindent The sums in the last two lines equal one, so $\mathcal{B} = 0$.
\end{proof}

%-----------------
\begin{proof}[Proof of \autoref{tnci}]
  Let $\bm{\delta}_i$ be as given in \autoref{adeg}, 
  \begin{align*}
    &\bm{d}_{-\N(i,K)}^L = \argmin\{ Y_i(\bm{\delta_i}, \bm{d}_{-\N(i,K)})\colon \bm{d}_{-\N(i,K)} \in \{0,1\}^{n-\abs{\N(i,K)}} \}, \quad\text{and} \\
    &\bm{d}_{-\N(i,K)}^U = \argmax\{ Y_i(\bm{\delta_i}, \bm{d}_{-\N(i,K)})\colon \bm{d}_{-\N(i,K)} \in \{0,1\}^{n-\abs{\N(i,K)}} \}.
  \end{align*}

  \noindent Then
  \begin{multline*}
    \abs{\mathcal{B}} \leq \frac{1}{n} \sum_{i=1}^n \big| \E\big[Y_i(\bm{\delta}_i,\bm{D}_{-\N(i,K)}) \mid T_i=t, \C_i\big] - \E\big[Y_i(\bm{\delta}_i,\bm{D}_{-\N(i,K)}) \mid T_i=t', \C_i\big] \big| \\
    \leq \frac{1}{n} \sum_{i=1}^n \big( Y_i(\bm{\delta}_i, \bm{d}_{-\N(i,K)}^U) - Y_i(\bm{\delta}_i, \bm{d}_{-\N(i,K)}^L) \big) \leq \Psi_K.
  \end{multline*}

  \noindent Consider the case where the unit-level effects are $\geq (\leq) \,0$. Then using \autoref{anp}, $\abs{\tau^*(t,t')} = (-)\tau^*(t,t') \geq \Delta_K > \Psi_K \geq \abs{\mathcal{B}}$. By \autoref{tdeg}, $\tau(t,t') = \tau^*(t,t') + \mathcal{B}$, so given non-negative unit-level effects, the right-hand side is non-negative.
\end{proof}

%-----------------
\begin{proof}[Proof of \autoref{tani}]
  By \autoref{aunc},
  \begin{multline}
    \tau(t,t') = \frac{1}{n} \sum_{i=1}^n \bigg( \sum_{\bm{d} \in \{0,1\}^n} \E[Y_i(\bm{d}) \mid \C_i] \prob(\bm{D}=\bm{d} \mid T_i=t, \C_i) \\ - \sum_{\bm{d}' \in \{0,1\}^n} \E[Y_i(\bm{d}') \mid \C_i] \prob(\bm{D}=\bm{d}' \mid T_i=t', \C_i) \bigg). \label{tas1}
  \end{multline}

  \noindent By definition,
  \begin{multline*}
    \E[Y_i(\bm{d}') \mid \C_i] = \E[\tilde{Y}_i^K(\bm{d}'_{\N(i,K)}) \mid \C_i] \\ + \E[g_n(i, \bm{d}', \bm{X}, \bm{A}, \bm{\varepsilon}) - g_{\abs{\N(i,K)}}(i, \bm{d}'_{\N(i,K)}, \bm{X}_{\N(i,K)}, \bm{A}_{\N(i,K)}, \bm{\varepsilon}_{\N(i,K)}) \mid \C_i].
  \end{multline*}

  \noindent Call the second term $\mathcal{B}_i(\bm{d}')$. Applying this identity twice and using \eqref{adegweak},
  \begin{multline*}
    \eqref{tas1} = \frac{1}{n} \sum_{i=1}^n \bigg( \sum_{\bm{d} \in \{0,1\}^n} \E[\tilde{Y}_i^K(\bm{d}_{\N(i,K)}) \mid \C_i] \prob(\bm{D}=\bm{d} \mid T_i=t, \C_i) \\ - \sum_{\bm{d}' \in \{0,1\}^n} \E[\tilde{Y}_i^K(\bm{\delta}_i) \mid \C_i] \prob(\bm{D}=\bm{d}' \mid T_i=t', \C_i) \bigg) + \frac{1}{n} \sum_{i=1}^n (\mathcal{B}_i(\bm{d})+\mathcal{B}_i(\bm{d}')).
  \end{multline*}

  \noindent The first term on the right-hand side reduces to 
  \begin{equation*}
    \frac{1}{n} \sum_{i=1}^n \sum_{\bm{d}_{\N(i,K)}} \big( \E[\tilde{Y}_i^K(\bm{d}_{\N(i,K)}) \mid \C_i] - \E[\tilde{Y}_i^K(\bm{\delta}_i) \mid \C_i] \big) \prob(\bm{D}_{\N(i,K)}=\bm{d}_{\N(i,K)} \mid T_i=t, \C_i),
  \end{equation*}

  \noindent which is $\tau^*_K(t,t')$ by \autoref{aunc}. Lastly, by \autoref{aunc}, \eqref{confounders}, and \autoref{aani}, $\mathcal{B}_i(\bm{d})+\mathcal{B}_i(\bm{d}')$ is uniformly bounded in absolute value by $2\gamma(K)$.
\end{proof}

%----------------------------------------------------------------------

\FloatBarrier
\phantomsection
\addcontentsline{toc}{section}{References}
\bibliography{exposed}{} 
\bibliographystyle{aer}

%----------------------------------------------------------------------

\end{document}